\documentstyle[aps,twocolumn,prl,epsf]{revtex}

\begin{document}
\draft

 \twocolumn[\hsize\textwidth\columnwidth\hsize\csname @twocolumnfalse\endcsname
\title{Non Fermi Liquid Dynamics of the Two-Channel Kondo Lattice}
\author{
M.\ Jarrell$^{(a)}$, Hanbin Pang$^{(a)}$, D.L.\ Cox$^{(b)}$, 
F.\ Anders$^{(b)}$,  and A.\ Chattopadhyay$^{(a)}$.
}
\address{
$^{(a)}$ Department of Physics,
University of Cincinnati, Cincinnati, OH 45221\\
$^{(b)}$Department of Physics,
The Ohio State University, Columbus, OH, 43202\\
}
\date{\today}
\maketitle

\widetext
\begin{abstract}
\noindent
The paramagnetic phase of the two-channel Kondo lattice model is examined 
with a Quantum Monte Carlo simulation in the limit of infinite dimensions.   
We find non-Fermi-liquid behavior at low temperatures including a finite 
low-temperature single-particle scattering rate, no Fermi 
distribution discontinuity, and zero Drude weight.  Both the optical
and quasiparticle mass enhancement and scattering relaxation rate show
consistent evidence of non-fermi liquid behavior.  However,
the low-energy density of electronic states is finite.
\end{abstract}
\pacs{Keywords:  non-Fermi liquid, Kondo lattice, dynamics}

 ] 

\narrowtext
\paragraph*{Introduction.}
The Fermi Liquid theory of Landau has provided a remarkably robust
paradigm for describing the properties of interacting Fermion systems 
such as liquid $^3$He and alkali metals (e.g., sodium). The key
notion of this theory is that the low lying excitations of the
interacting system possess a 1:1 map to those of the noninteracting
system and hence are called ``quasiparticles.''  In the metallic
context, one may think of the quasiparticles as propagating electron-like 
wave packets with renormalized magnetic moment and effective mass
reflecting the ``molecular field'' of the surrounding medium.
A sharp Fermi surface remains in the electron occupancy function
$n_{\vec k}$ which measures the number of electrons with a given
momentum, and for energies $\omega$ and temperatures $T$
asymptotically close to the Fermi surface the excitations have a
decay rate going as $\omega^2 + \pi^2 (k_BT)^2$, which is much smaller
than the quasiparticle energy.  This generally translates into a $T^2$
contribution to the electrical resistivity $\rho(T)$ and an $\omega^2$
contribution to the scattering relaxation rate $\Gamma(\omega)$ inferred
from the optical conductivity, and both the optical and 
single-particle electronic masses are found to be enhanced but positive
and finite.  
   
The Fermi liquid paradigm appears now to be breaking down empirically
in numerous materials, notably a number of fully three dimensional
heavy Fermion alloys and compounds\cite{hfnfl}.  In these systems such 
anomalies as a conductivity with linear dependence on $\omega,T$ and
logarithmically divergent linear specific heat coefficients are often
observed.  In addition, some heavy fermion lattice compounds, such
UBe$_{13}$\cite{ottold} and CeCu$_2$Si$_2$\cite{steggoa}, the 
superconductivity arises in a normal state which is clearly {\it not} described 
as a Fermi liquid.  Specifically, the resistivity is approximately linear in $T$ 
above the transition, and the magnitude of the resistivity at the transition is 
high (typically 80-100 $\mu-\Omega$-cm in UBe$_{13}$ even in the best samples). 
If the quasiparticle paradigm indeed breaks down, this may require completely
new concepts to explain the superconducting phases of these materials.
While the Luttinger liquid theory provides an elegant way to achieve 
non-Fermi liquid theory in one-dimension (with, e.g., no jump discontinuity 
in $n_{\vec k}$, and separation or unbinding of spin and charge quantum 
numbers), this results from the special point character of the Fermi surface. 
Whether the essential spin-charge separation may ``bootstrap'' into higher 
dimensions remains unclear\cite{lutnfl}.  
Among the remaining theories to 
explain experiment are those based upon proximity to a zero temperature 
quantum critical point\cite{qcpnfl}, those based upon disorder induced 
distributions of Kondo scales in local moment systems\cite{disknfl}, and those 
which hope to explain the physics from impurity to lattice crossover effects 
in the multi-channel Kondo model\cite{twochnfl}.  Notably lacking for 
dimensions higher than one are rigorous solutions to microscopic models 
which display non-Fermi liquid behavior\cite{butkotsi}.
 
In this paper, we present the first rigorous solution for the dynamics of the
two-channel Kondo lattice model in infinite spatial dimensions.  We
find that the paramagnetic phase of this model is an ``incoherent metal''
with finite density of states at the Fermi energy and finite residual
resistivity.  Both the single-particle and the optical excitations
are non-Fermi liquid like; in particular, there is a finite lifetime for 
electrons at the Fermi energy, an ill defined quasiparticle mass, a linear 
low temperature electrical resistivity with a finite residual value, 
a negative optical mass and a linear in frequency low temperature
scattering relaxation rate.  
  
\paragraph*{Motivation} The two-channel Kondo impurity model consists of
two identical species of non-interacting electrons antiferromagnetically
coupled to a spin 1/2 impurity.  Non-Fermi liquid behavior results
because of the inability to screen out the impurity spin:  it is
energetically favorable for both conduction electron bands to couple to 
the impurity which gives a spin 1/2 complex on all length scales as the 
temperature goes to zero.  As a result, the ground state is degenerate and 
the excitation spectrum non-Fermi liquid like.  We believe that at least 
the quantitative features of the model will change once the concentration 
of impurities becomes finite.  A single impurity always lives in a Fermi liquid 
host; whereas the introduction of a finite concentration of impurities into 
the system will mean that at some temperature, the host of each impurity will 
be a non Fermi liquid due to the presence of the others.  The simplest model 
to test this hypothesis is the fully concentrated one; ie. the two-channel 
Kondo lattice model.

\paragraph*{Model} The Hamiltonian for the two-channel Kondo
lattice is 
$$H=J\sum_{i,\alpha}{\bf{S}}_i\cdot {\bf{s}}_{i,\alpha}
    -{t^*\over 2\sqrt{d}}\sum_{<ij>,\alpha,\sigma}
    \left(c_{i,\alpha,\sigma}^{\dag} c_{j,\alpha,\sigma}
    +\mbox{h.c.}\right) $$
\begin{equation}
    -\mu\sum_{i,\alpha,\sigma}c_{i,\alpha,\sigma}^{\dag}
    c_{i,\alpha,\sigma}\,,
    \label{Ham}
\end{equation}
where $c_{i,\alpha,\sigma}^{\dag}$ ($c_{i,\alpha,\sigma}$)
creates (destroys) an electron on site $i$ in channel $\alpha=1,2$
of spin $\sigma$, ${\bf{S}}_i$ is the Kondo spin on site $i$, and
${\bf{s}}_{i,\alpha}$ are the conduction electron spin operators
for site $i$ and channel $\alpha$.  The sites $i$ form an
infinite-dimensional
hypercubic lattice.  Hopping is limited to nearest neighbors with
hopping integral $t\equiv t^*/2\sqrt{d}$; 
the scaled hopping integral $t^*$ determines the energy unit and is set
equal to one $(t^*=1)$.  Thus, on each site the Kondo spin mediates spin
interaction between the two different channels.  This problem is non
trivial, and for the region of interest in which $J>0$ and $T\ll J$,
$t^*$ it
is describable only with 
non-perturbative approaches.  
Clearly some simplifying method which allows for a
solution of the lattice problem in a non-trivial limit is necessary.

\paragraph*{Formalism and Simulation}
Such a method was proposed by Metzner and Vollhardt \cite{mevoll} who
observed that the renormalizations due to local two-particle interactions
become purely local as the coordination number of the lattice increases.
A  consequence is that the solution of most standard lattice
models may be mapped onto the solution of a local correlated system coupled
to an effective bath that is self-consistently
determined\cite{bramiel90_to_georges92}.  We 
refer the reader to the above references and recent reviews for further
details on the method\cite{infdrev}.  
In order to solve the remaining impurity problem, we use the Kondo 
impurity algorithm of Fye and Hirsch\cite{fye}, modified to simulate the
two-channel problem\cite{luk}.  
The results presented here are limited to the model at half filling of the
conduction bands ($N=1.0$ for $J=0.75, 0.625, 0.5, 0.4$).  A sign problem 
was also encountered in the QMC process which limited our access to very low 
temperatures.  The Euclidean-time QMC results for the local greens function 
$G(\tau)$ were then analytically continued to real frequencies using the 
``annealing'' Maximum Entropy method\cite{jarrell_mem}.  
The single particle self energy
may then be obtained by inverting the relation 
$G(\omega)=-i\sqrt{\pi} w\left(\omega+\mu-\Sigma(\omega)\right)$,
where $w(z)$ is the complex Fadeev function.  
Error bars are ill-defined
for individual points in analytically continued spectra\cite{jarrell_mem},
and are less than 3\% for the other results presented here.
  
\paragraph*{Results.}  When the temperature is lowered below the lattice Kondo
temperature $T_0$\cite{jarrell_2ckl}, we find non-Fermi liquid behavior in the 
single-particle properties of the model.  For example, the derivative of the 
particle distribution function $d n(\epsilon_k)/d \epsilon_k 
=-T\sum_n1/(i\omega_n-\epsilon_k+\mu-\Sigma(i\omega_n))^2$ saturates to a 
finite width distribution\cite{jarrell_2ckl}, in contrast to a Fermi liquid
where it would display a dominant delta function contribution as $T\to 0$, 
and a Luttinger liquid\cite{lutnfl} or Marginal Fermi liquid\cite{ffnj} 
where it would have a singular divergence.    

In Fig. 1 we show one electron properties of the model.  Fig.~1(a) displays the 
single particle DOS.  This has a finite value as $\omega,T\to 0$, with a peak 
away from the Fermi energy.  Novel behavior is seen in the real part of the 
one electron self energy which has positive slope at $\omega\to 0$ (Fig.~1(c)).  For a Fermi liquid, this slope would be negative.  The physical content is
important: $Z=1/(1-\partial Re\Sigma/\partial \omega)$ measures the
overlap of the quasiparticle wave function with the original
one-electron wave function having the same quantum labels. A positive
slope leads to $Z>1$ or $Z<0$, indicating a breakdown of the quasiparticle
concept.  Concomitant with the finite width of $-dn/d\epsilon_k$ is a 
finite imaginary part to the low temperature self energy (Fig.~1(b)).  
This indicates that the one-electron excitations are ill defined on approach 
to the Fermi surface, again ruling out a Fermi liquid description.   Since 
the low temperature thermodynamic properties such as the specific heat, 
uniform magnetic susceptibility, and charge susceptibility display no evidence 
for a gap, we believe the observed behavior indicate a new kind of metallic 
state.  

From our numerical results, one can show that the zero temperature self energy 
must be non-analytic.  If the self energy is analytic everywhere in the 
upper half complex plane (and infinitesimal close to the real axis), then 
one may easily show that
\begin{equation}
\lim_{T\to 0} \frac{{\rm{Im}}\Sigma(i\omega_0)}{\omega_0}=
\lim_{T\to 0}
\left.\frac{d{\rm{Re}}\Sigma(\omega)}{d\omega}\right|_{\omega=0}\, ,
\label{analsig}
\end{equation}
where $\omega_0=\pi T$ is the lowest Fermionic Matsubara frequency
along the imaginary axis. However, we find that when $T<T_0$,
$\frac{{\rm{Im}}\Sigma(\omega_0)}{\omega_0}<0$ and is apparently divergent;
whereas, from Fig.~2 it is apparent that
$\left.\frac{d{\rm{Re}}\Sigma(\omega)}{d\omega}\right|_{\omega=0} \approx 1.5$.
Thus, $\Sigma(\omega)$ cannot be analytic at the origin of the complex
plane.  
 
The non-Fermi liquid behavior also strongly effects experimentally relevant 
transport properties of the system.  The electrical resistivity is shown in 
Fig.~2.  We find that $\rho(T)/\rho(0)$ curves for different $J$ values
collapse onto a universal scaling curve when plotted against $T/T_0$.  As 
shown in the inset, $\rho(T)\simeq\rho(0)[1+B(T/T_0)]$ for $T\to 0$, with 
$B<0$.  We interpret the finite value of $\rho(0)$  
together with $-Im\Sigma(0,0)$ as ``spin disorder scattering'' off of 
the degenerate screening clouds centered about each local moment spin.  Given 
the finite density of one particle excitations at the Fermi energy, this 
finite residual resistivity is indicative of an ``incoherent metal phase'' 
brought about by the disordered spin degrees of freedom, in qualitative 
agreement with results obtained with a Lorentzian bare conduction DOS 
(which doesn't self-consistently renormalize)\cite{coxinfd}.  
This surprising fact of a finite resistivity for $T\rightarrow 0$ in a
translational invariant system reflects the infinitely degenerate
ground state of the two channel lattice with a residual entropy.

The optical conductivity $\sigma_1(\omega)$ is measured in units of 
$\sigma_0=e^2\pi/2\hbar a$, which varies between 
$10^{-3}...10^{-2}[(\mu\Omega cm)^{-1}]$, depending on the 
lattice constant $a$. As shown in Fig.~3(a), consistent with the lack of 
quasiparticles, the low temperature optical conductivity displays vanishing 
Drude weight $D$.  To confirm this, we also calculated $D$ by 
extrapolation\cite{swz}, 
$D\propto \chi_{jj}(q=0,i\omega_\alpha \to0)-\chi_{jj}(q=0,i\omega_\alpha =0)$,
where $\chi_{jj}(q=0,i\omega_\alpha)$ is the bulk current-current 
susceptibility.  For each value of $J$ and all fillings, we found that the 
extrapolated zero-temperature Drude weight was zero.   In addition, 
$\sigma_1(\omega)$ has a finite-frequency ($\omega\approx0.6J$) peak.  Both 
these features again support our interpretation in terms of a new kind of
non-Fermi liquid metallic state.

	The optical conductivity of metals, even non-Fermi liquid 
metals\cite{degiorgi}, is usually analyzed by rewriting it in a 
generalized Drude form
\begin{equation}
\sigma(\omega)= \frac{\omega_p^2}{4\pi} \frac{1}{\Gamma(\omega) 
-i\omega \left(1+\lambda(\omega)\right)}\,,
\label{gen_drude}
\end{equation}
where $\sigma(\omega) = \sigma_1(\omega) + i \sigma_2(\omega)$.  The resulting 
mass enhancement $\left(1+\lambda(\omega)\right)$ and scattering relaxation rate
$\Gamma(\omega)$ are shown in Fig.~4 when $J=0.625$ for various temperatures.  The lifetime approximation becomes exact in infinite dimensions, thus 
at low $\omega$ one expects 
$1+\lambda(\omega)=1-\frac{2}{\omega}{\rm{Re}}\Sigma(\omega/2)$
and 
$\Gamma(\omega)=-2{\rm{Im}}\Sigma(\omega/2)$\cite{coxandgrewe}.  At high 
temperatures, $\Gamma(\omega)$ and $\left(1+\lambda(\omega)\right)$ are essentially 
flat so a Drude peak is recovered in $\sigma_1(\omega)$; whereas at lower
temperatures they become strongly frequency dependent.  Here, the behavior 
of $\Gamma(\omega)$ is consistent with the imaginary part of the self 
energy, indicating that the optical and quasiparticle scattering rates are 
consistent for this model, even though quasiparticles do not exist.  
Note that when $T/T_0 =0.20 $, the low frequency optical mass is negative,
$m^*(0)\approx-1.0$.  This behavior is only qualitatively consistent with the 
mass enhancement estimated from the low temperature self energy $1-\left.d{\rm{Re}}\Sigma(\omega)/d\omega\right|_{\omega=0}\approx -0.5$
(cf.\ Fig.~1).  For larger values of $J>0.75$, $m^*(0)\agt0$.  Thus, the
negative mass enhancement should be viewed as a sufficient but not
necessary identifying feature of a non Fermi Liquid.

Finally, we mention the possible applicability of our results to
concentrated heavy Fermion systems.  Three systems display resistivity 
of the form $\rho(T)\approx \rho(0)[1+B(T/T_0)^{\alpha}]$ for $T<T_0$, 
with $\rho(0)$ of order the unitarity limit and $\alpha \approx 1$.
They are: UCu$_{5-x}$Pd$_x$ (with $B<0$)\cite{hfnfl}, UBe$_{13}$ 
($B>0$)\cite{ottold,aronson,stegnew}, and CeCu$_2$Si$_2$ 
($B>0$)\cite{steggoa}.  UBe$_{13}$ and CeCu$_2$Si$_2$ are ordered 
compounds which have been proposed as possible two-channel lattice 
systems (see Refs.\cite{twochnfl}(b,c)), while UCu$_{5-x}$Pd$_x$ is a 
possible example of the distribution of Kondo scales scenario\cite{disknfl}. 
In both UBe$_{13}$\cite{bonn} and UCu$_{5-x}$Pd$_x$\cite{degiorgi} the
measured $\sigma_1(\omega)$ and $\Gamma(\omega)$ are strikingly similar 
to those displayed in Figs.~3 and 4.

We would like to acknowledge useful discussions with W.\ Chung,
A.\ Georges, M.\ Ma, A.J.\ Millis, and W.\ Putikka.  We especially thank
L.\ Degiorgi for a thoughtful discussion of these results.
Jarrell and Pang would like to acknowledge the support of NSF grants 
DMR-9406678 and DMR-9357199.  Cox acknowledges the support of the U.S. 
Department of Energy, Division of Basic Energy Sciences, Office of 
Materials Research, and, at the ITP, by NSF Grant No. PHY94-07194. 
Anders was funded by the Deutsche Forschungsgemeinschaft under An 275/1-1.
Computer support was provided by the Ohio Supercomputer Center.

\begin{figure}[t]
\epsfxsize=3.8in
\epsffile{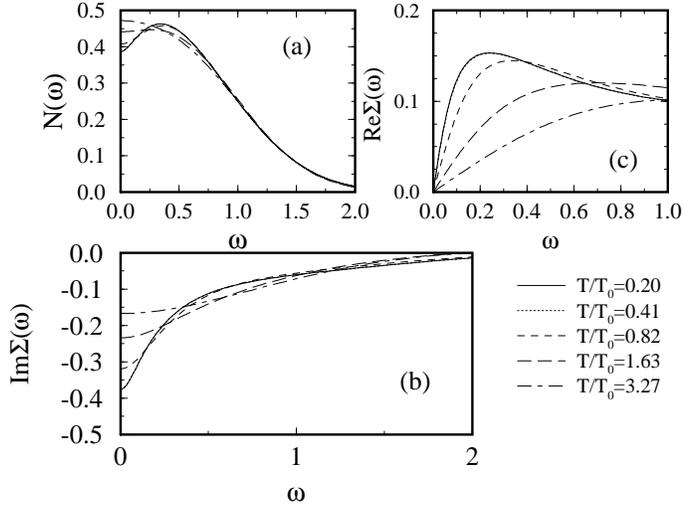}
\caption{Single particle properties of the two-channel Kondo lattice in
infinite dimensions when $J=0.625$ and $N=1.0$.  (a) Single-particle density 
of states (DOS).  At high temperatures $T\gg T_0$ (not shown), the DOS is 
a Gaussian, crossing over to the peaked distribution with relative suppression  
at $\omega\to 0$ for lower temperatures when $T\ll T_0$.  (b) Imaginary part 
of the self energy.  As the temperature is lowered the self energy does not 
approach a Fermi liquid form ${\rm{Im}}\Sigma(\omega)\propto -T^2-\omega^2$, 
but rather appears to approach the non-analytic form (see text)
${\rm{Im}}\Sigma(\omega)\propto -c+\left|\omega\right|$.  (c) The real
part of the self energy ${\rm{Re}}\Sigma(\omega)$, is also anomalous since 
its initial slope is positive indicating a quasiparticle renormalization 
factor which is greater than one.}

\end{figure}

\begin{figure}[htb]
\epsfxsize=3.8in
\epsffile{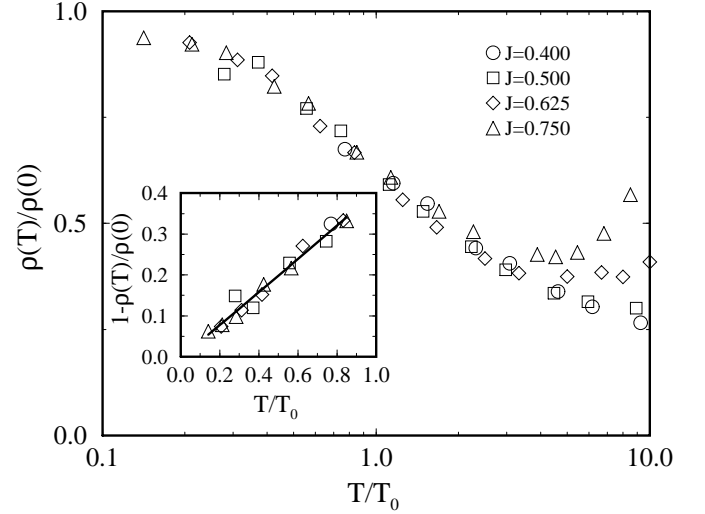}
\caption{Resistivity of the two-channel Kondo lattice.  $\rho(T)/\rho(0)$ is 
plotted versus $T/T_0$ for various values of $J$.  In the inset, the lowest 
temperature data (for $T/T_0 < 1$, shown as open circles) was fit 
to $\rho(T)/\rho(0)=1+B\left( T/T_0\right)^\alpha$, with $B=-0.4$ and 
$\alpha=1.03$. }
\end{figure}

\begin{figure}[htb]
\epsfxsize=3.8in
\epsffile{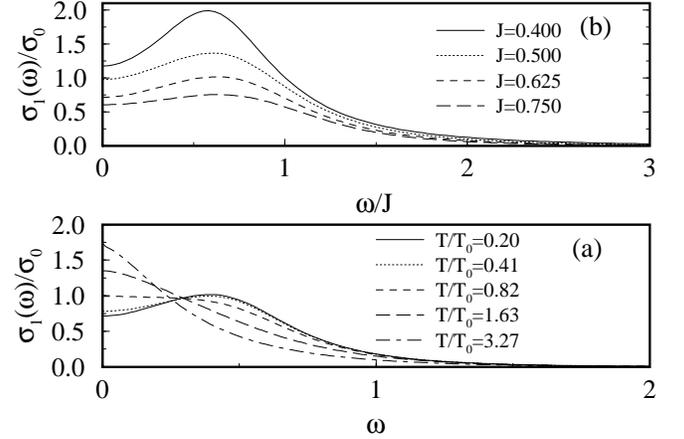}
\caption[]{{Optical conductivity of the two-channel Kondo lattice.  (a) 
Optical conductivity when $J=0.625$.  As the temperature is lowered $T<T_0$, 
the optical conductivity develops a pseudogap at low frequencies.  No
evidence of a finite Drude weight, $D$, can be seen, consistent with estimates
of $D$ obtained from extrapolation of Matsubara-frequency results\cite{swz}.
(b) The optical conductivity vs.\ $\omega/J$ when $T=0.0156$ for various 
values of $J$.  $\sigma_1(\omega)$ displays a finite-frequency peak
at roughly $\omega\approx 0.6J$.}}
\end{figure}

\begin{figure}[htb]
\epsfxsize=3.8in
\epsffile{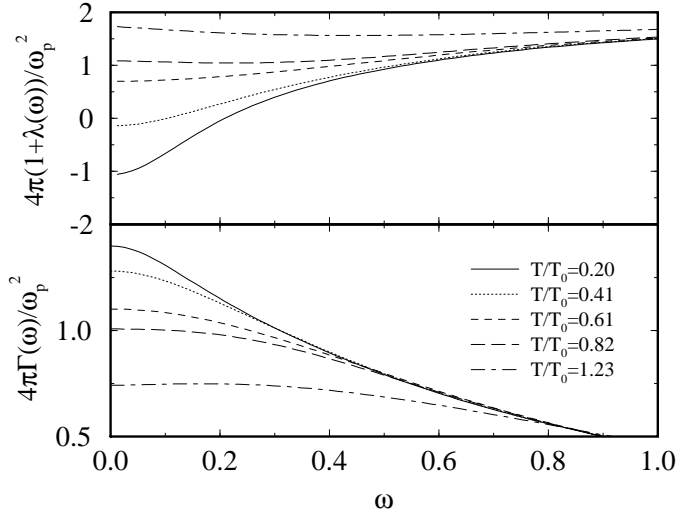}
\caption[]{{Optical scattering rate $\Gamma(\omega)$(a) and mass enhancement
$\left(1+\lambda(\omega)\right)$ (b) obtained from the optical conductivity shown in Fig.~2(a)
using Eq.~\ref{gen_drude}.  For $T/T_0 \ll 1$ both $\Gamma(\omega)$ and
$\left(1+\lambda(\omega)\right)$ display non fermi liquid behavior:  $\Gamma(\omega)$ is
roughly linear in $\omega$ and $m^*(0)<0$.  Typical Fermi liquid behavior
is recovered when $T/T_0 \agt 1$, since both $\Gamma(\omega)$ and
$\left(1+\lambda(\omega)\right)$ become weakly frequency dependent consistent with the recovery
of the Drude peak in Fig.~3(a).}}
\end{figure}

\end{document}